# Numerical experiments with assimilation of the mean and unresolved meteorological conditions into large-eddy simulation model

Igor Esau[1]*

* G.C. Rieber Climate Institute of the Nansen Environmental and Remote Sensing Center, Bergen, Norway

**Abstract.** Micrometeorology, city comfort, land use management and air quality monitoring increasingly become important environmental issues. To serve the needs, meteorology needs to achieve a serious advance in representation and forecast on micro-scales (meters to 100 km) called "meteorological terra incognita". There is a suitable numerical tool, namely, the large-eddy simulation modelling (LES) to support the development. However, at present, the LES is of limited utility for applications as it cannot start from (i) scarcely measured atmospheric conditions and (ii) it cannot account for unresolved surface details. This study presents an analysis of several numerical experiments with the LESNIC LES code. The experiments were aimed to test the prospective ways to improve the LES utility for the applied problems. The study addresses two problems. First, the data assimilation problem on micro-scales is investigated as a possibility to recover the turbulent fields consistent with the mean meteorological profiles. Second, the methods to incorporate of the unresolved surface structures are investigated in *a priopi* numerical experiments. The numerical experiments demonstrated that the simplest nudging or Newtonian relaxation technique for the data assimilation is applicable on the turbulence scales. It is also shown that the filtering property of the three layers' artificial neural network (ANN) can be used for formulation of the surface stress from the unresolved surface features. Introduction of independently trained ANN for each of dynamical sub-regions in the LES domain could greatly reduce computer time needed to estimate closure coefficients through omitting multi-layer explicit filtering in the dynamic closure. Moreover, the ANN is shown to be a robust predictor for scalar concentrations in the urban sub-layer with unresolved scalar sources.

## 1. Introduction

Meteorology has many applications beyond the traditional weather forecast or climate change studies. Several applications, e.g. city comfort, land use management and air quality monitoring, require development of micro-meteorological methods. To serve these applications, micro-meteorology needs to achieve a serious advance in representation and forecast on micro-scales (meters to 100 km) called "meteorological terra incognita" by J. Wyngaard [1]. Given the prospects for computational resources, it is reasonable to expect meteorological modelling to be feasible on scales 100 m or so in the nearest future. On these scales, we are in the realm of highly turbulent flows. Moreover, the turbulence, at least in the eddy generation interval of scales [2], has to be resolved. There is a suitable turbulence-

---

resolving numerical tool, namely, the large-eddy simulation (LES) models available. In this study, we use the LES code LESNIC developed at Uppsala University and the Nansen Environmental and Remote Sensing Center. The code details can be found in Esau [3]. The LES technique has been known in meteorology for more than 40 years. Nevertheless, the limited computational resources do not allow its application in the practical meteorology. The LES field was restricted by case studies predominantly over homogeneous surfaces. More interesting applications like urban "hot-spots" of anthropogenic impact on the Earth's system were not fully addressed. At present, the LES applications are limited by (i) the absence of assimilation techniques to start from scarcely measured atmospheric conditions and by (ii) its inability to incorporate unresolved surface details.

This study addresses two possible approaches to deal with both issues. The data assimilation problem on meteorological micro-scales is investigated using the simplest nudging or Newtonian relaxation technique. We study the Ekman boundary layer as the most theoretically understood case in the boundary layer meteorology. The LES experiments aim to investigate a possibility to recover the turbulent fields consistent with the mean meteorological profiles. Generally speaking, the problem may not have any or may have non-unique solution. Nevertheless, a part of the problem, namely, reconstruction of the turbulent statistics, typically in vertical direction, consistent with the given mean profiles is nothing else but the widely employed turbulence closure approach. The idea of nudging is sufficiently simple to attract attention of LES modellers. Porson et al. [4] used nudging in LES to simulate observed nocturnal fog cases. Porson et al. (personal communication) noticed however a certain lack of fluctuations in their runs, probably because of too small relaxation time scale of just 900 s. Cheng et al. [5] used nudging for the averaged velocity field and several micro-physical parameters only on 5 outermost points in (periodic) horizontal plain so that perturbations flow through the nudged area in simulations. But the authors did not investigate the effect of nudging on their simulations. The nudging of micro-physical parameters in the LES study was also demonstrated by Carrio et al. [6].

The unresolved surface structures can be incorporated in three ways: through the traditional parameterization methodology, which has demonstrated its weakness in the case of a few elements dominate the unresolved fluxes; through adjoint optimisation model [7,8]; and through artificial neural network technique (ANN). The latter looks very attractive due to its generality, computational efficiency and robustness in the case of strong patchiness of the surface features. Here, we investigate applicability and utility of the three layers' artificial neural network in *a priopi* numerical experiments. ANN is a statistical technique, which solves a set of parametric function equations, called static nonlinear maps, with empirically tuned coefficients or weights to relate the model variables to a unresolved quantity of interest. The weights are obtained in so-called training process or a search for the global minimum of a cost function in multi-dimensional parameter space. White [9] proved that any nonlinear function could be approximated with any degree of accuracy and unit probability using sufficiently large weight matrices. ANN were found useful in different applications, e.g. turbulence control [10,11], turbulent flow simulations [12,13], air quality monitoring [14,15,16,17].

The paper is structured as follows. Section 2 presents the analysis of the data assimilation experiments with the LES. Section 3 presents the analysis of the ANN application to the LES data. Section 4 outlines the conclusions.

## 2. Data assimilation into large-eddy simulation experiments

*a. The method*

In the planetary boundary layer, the turbulence plays an essential role. Therefore the turbulent stress (diffusion), induced by meteorological fluctuations on scales between mm to km, cannot be omitted in the problem formulation. The assimilation problem is to reconstruct statistically consistent fields of fluctuations to match the averaged observed profiles of the meteorological quantities. In this formulation, the large-scale (mean) component of the meteorological fields is to be relaxed to the observed values but the-scale (fluctuating) component is to be simulated in the statistical equilibrium with the large scale component.
The nudging is based on the simple idea of the Newtonian relaxation. The model equations are supplement with a new term in the right hand side, proportional to the difference between the calculated and observed values. The nudging can be interpreted as a variant of the Kalman-Bucy filter (a continuous time version of the Kalman filter) with the gain matrix prescribed rather than obtained from co-variances. In its general form, the idea has been formulated by von Storch et al. [18] as the spectral nudging in the limited area models as

$$\Psi(x,y,z,t) = \sum_{n=-N_x; m=-N_y}^{N_x; N_y} \alpha_{nm}(z,t) \exp(inx/L_x) \exp(imy/L_y). \tag{1}$$

$\Psi(x,y,z,t)$ is any model variable and $\alpha_{nm}(z,t)$ are its Fourier transformation coefficients. Similar decomposition can be used for observed quantities as

$$\Psi^{obs}(x,y,z,t) = \sum_{n=-M_x; m=-M_y}^{M_x; M_y} \beta_{nm}(z,t) \exp(inx/L_x) \exp(imy/L_y) \tag{2}$$

In our problem, $\beta_{nm}(z,t)$ are the Fourier transformation coefficients. The number of Fourier coefficients in observations is always less than in a model, i.e. $N_x > M_x$ and $N_y > M_y$. The nudging is achieved through the following modification of tendencies

$$\frac{\partial \Psi}{\partial t} = R + t_r^{-1}(\Psi^{obs} - \Psi) = R + \sum_{n=-M_x; m=-M_y}^{M_x; M_y} \eta_{nm}(z)[\beta_{nm}(z,t) - \alpha_{nm}(z,t)]e^{inx/L_x + imy/L_y}, \tag{3}$$

where $R$ is the terms in the right hand side. The confidence coefficients $\eta_{nm}(z)$, which weight the model against the observations, are chosen in a way to relax the largest modes but to leave the shorter, turbulence modes. Usually only just a few largest scale $\eta_{nm}(z)$, e.g. $n = 0 \ldots 3$ and $m = 0 \ldots 5$, are set to non-zero values. Instead of sharp cut off of the coefficients, a smoother damping can be used (A. Glazunov, personal communication, December 2008). In the simplest version of nudging in Eq. (3) only one coefficient corresponding to the mean (horizontally averaged) value of the variables is used. This is the case of the present study.

*b. Results*

To study the described nudging in the LES, we adopted the well studied Andren et al. [19] case for the Ekman boundary layer. The results for simulation of the Andren case with

LESNIC model can be found in Esau [3]. This run is used as the control (CTR) simulations. Here we investigate the properties of the simple nudging (NDG) and its effect of the turbulence. The relaxation time scale varies from 600 s to 60000 s in different experiments. For the time scale less than about 3600 s (one hour), the damping effect has been found too strong. For the time scale larger than about 10800 s (3 hours), the nudging failed to suppress the inertial oscillation. Thus we will discuss the run with the relaxation time scale 1 hour.

Figure 1 shows the normalized difference, CTRL-NDG, between the streamwise turbulent kinetic energy spectra for each height in the simulation domain. As one can observe, the nudging, as expected, damps the largest frequencies, and hence spatial scales, at all heights. Unexpectedly it has two side effects. The first one is that it damps also all frequencies at the EBL top. Turbulence in the EBL bottom remains mostly unmodified but in the mid-EBL on moderate frequencies, the amplification is observed. The integrals (over the EBL sub-layers) of the energy spectra are presented in Figure 2.

In order to interpret the effect of nudging, one should notice that the observed profile includes all forces and transient effects, which are not accounted for in the model formulation. The initial profile in the Andren case was generated by a single-column turbulence model, which is not necessarily consistent with any of the physically observed cases. The damping in the EBL top reflects the fact that the single-column models are turned to conventionally neutral EBL without the recognition of the suppressing effect of the free atmosphere stratification. It has been clearly demonstrated in Mauritzen et al. [20]. The nudging must suppress the fluctuations as well. But then, in order to keep the consistency in the lower EBL, the model has to intensify the turbulence near the surface. This mechanism has been demonstrated in [3]. Thus, the model adequately recovers the turbulence fluctuations. The turbulence amplification in the mid-layer may have a different dynamical explanation. Starting from the Andren case without nudging, the LES models develop the inertial oscillation. The nudging damps the oscillation but in result, it also changes the instant curvature of the velocity profile and therefore the flow hydro-dynamic stability. The prescribed mean velocity profile is less stable (more concave) in the mid-EBL. This subtle modification of the curvature intensifies the turbulence.

Concluding, it is reasonable to claim that the nudging technique is a promising candidate for the data assimilation in the turbulence-resolving LES. The interesting results have been published already. Moreover, this study reveals no significant distortions of the turbulent spectra of the total turbulent energy of fluctuations on the essential scales of the turbulence generation and down to the dissipation. Even in the case of dynamical inconsistency of the observed profile, the LES tries to generate proper turbulent fluctuations through intensification of the turbulence in one or another sub-layer where the stability restrictions are relaxed. Probably more sophisticated procedure is needed to account for the second-order effects.

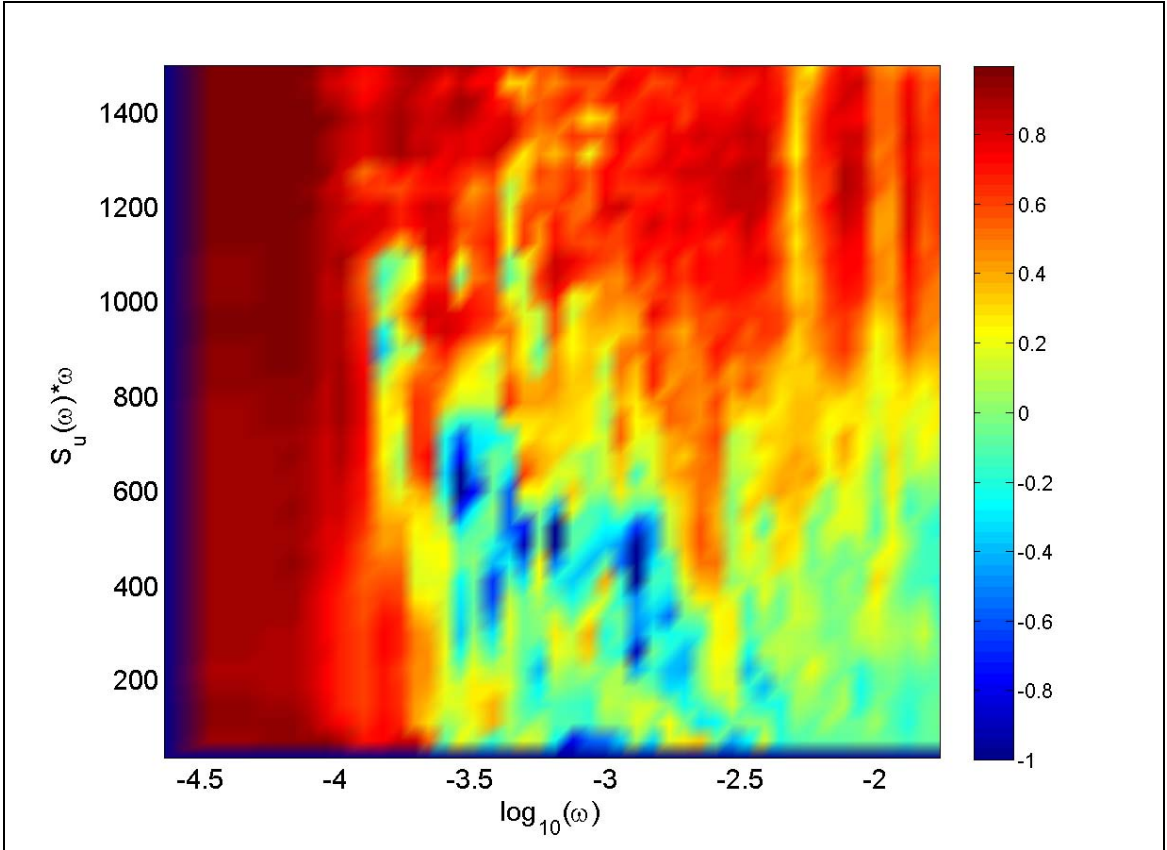

Figure 1. The normalized (+1 corresponds to the strongest damping and -1 corresponds to the strongest amplification) difference CTRL-NDG between the streamwise turbulent kinetic energy spectra as function of height in the Andren case [19] experiment with LESNIC model, $\omega$ is the frequency (Hz).

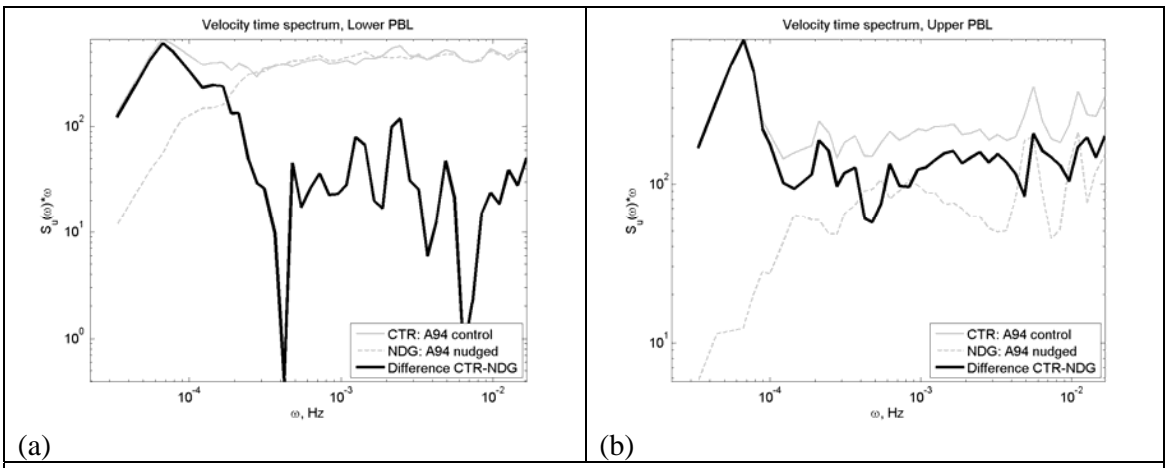

(a)  (b)

Figure 2. The spectra as in Figure 1 but integrated over lower PBL (0-800 m) in panel (a) and the upper PBL (800-1500 m) in panel (b).

## 3. Assimilation of unresolved surface features in large-eddy simulation experiments

*a. The idea*

Limitations on computational resources and a lack of knowledge to parameterize sub-grid scale processes in coarsely resolved simulations have motivated a number of attempts to make shortcuts between practically important but unresolved quantities, like concentrations in the urban sub-layer, and more readily simulated ones, like the mean wind.

The idea behind the use of the artificial neural network (ANN) is fairly simple. Suppose a certain feature in the boundary conditions cause a perturbation in the modelled variables. For instance, a building, unresolved with the given model mesh, causes perturbation of the flow. Then observing the persistent difference between the observed and model flows, one can introduce a statistical correction as an additional artificial force into the model. Since the structure of this force in typically quite complex, it is convenient to use a set of linear equations with empirical coefficients. The coefficients are chosen in such a way to minimize the difference between the model and the observations in some sense. Since there are many ways to define the sense of the optimization, multitude of the ANN methods were proposed.

Mathematically, the ANN can be considered as an optimal estimator of unresolved quantities using a finite time series of resolved-scale solutions [21]. The ANN could be constructed as a set of parametric function, which coefficients need to be found in so-called training process through the statistically fitting (or learning function) to the observed values. Specifically, the three-level ANN (one input layer of 16 neurons and two hidden layers of 4 an 1 neurons) with a back-propagation learning function and the tan-sigmoid transfer function. This configuration is similar to "ATMOSFERA" and Sarghini et al. [22] configurations. The neuron distribution was selected optimal by trail-and-fail approach for the particular type of testing data in this paper. The ANN could be written as

$$q(x_i) = \sum_m^M C_m \sum_{j=1}^N A_j \tanh\left(\sum_{k=1}^{N_k} B_{jk} X_m(x_k) + b_{jm}\right) + a_m ,  \qquad (4)$$

where $c_x$ and $X$ are the scalar output of the network and the vector input. Weights $A_j$, $B_{jk}$, $C_m$ and thresholds $b_{jm}$ and $a_m$ allow approximation of practically any analytical function with sufficiently large input time set. To obtain values of these coefficients, the network was trained using conjugate-gradient method for the cost minimization on the 5 min subset of the available LES data as described below. Actual calculations have been run in MATLAB using neural network toolbox (*nnet*) by Demuth and Beale.

Kukkonen et al. [16] evaluated performance of five ANN methods, a linear regression statistical model and a hydrodynamic model using measurements of urban chemical concentrations at two stations in central Helsinki from 1996 to 1999. On average, the ANN methods demonstrated better agreement with data than other types of models. Mammarella et al. [17] reported the use of the ANN in the fully operational air quality monitoring system "ATMOSFERA". It requires preconditioning through the model feedback, as large predictor vector dimension leads to a classical ill-conditioned problem of the linear algebra [23]. Schematically, the boundary condition problem is shown in Figure 3.

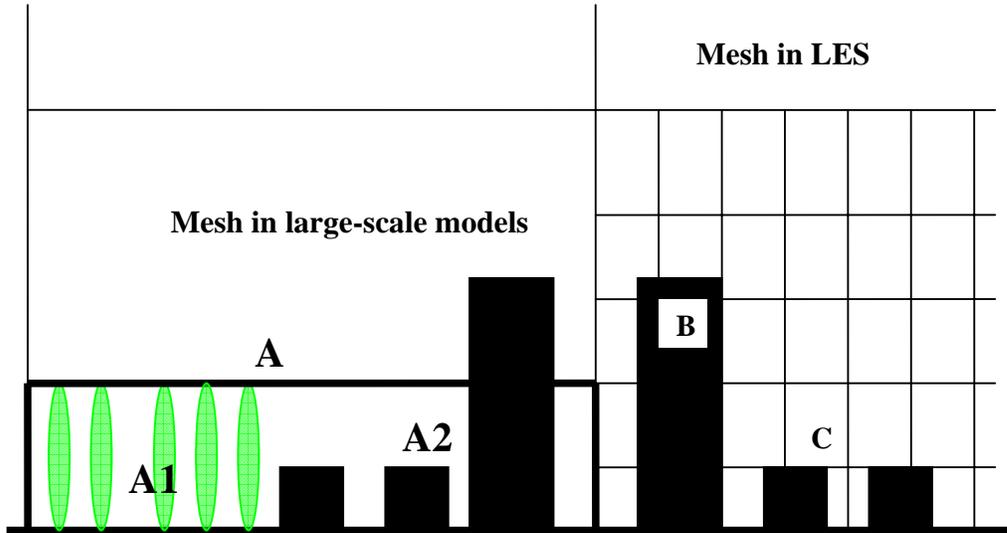

**Figure 3.** Schematic representation of the urban relief on large- and fine-scale model meshes. At any feasible mesh, the relief and surface features remain to be under-resolved as building to PBL scale ration 1:500 is currently can be resolved only in street canyon models but not on the megacity scales. **C** and **B** mark the roughness spot and the obstacle used in the numerical experiments with the ANN corrections. Runs LES-H1 and LES-H2 in Table 1.

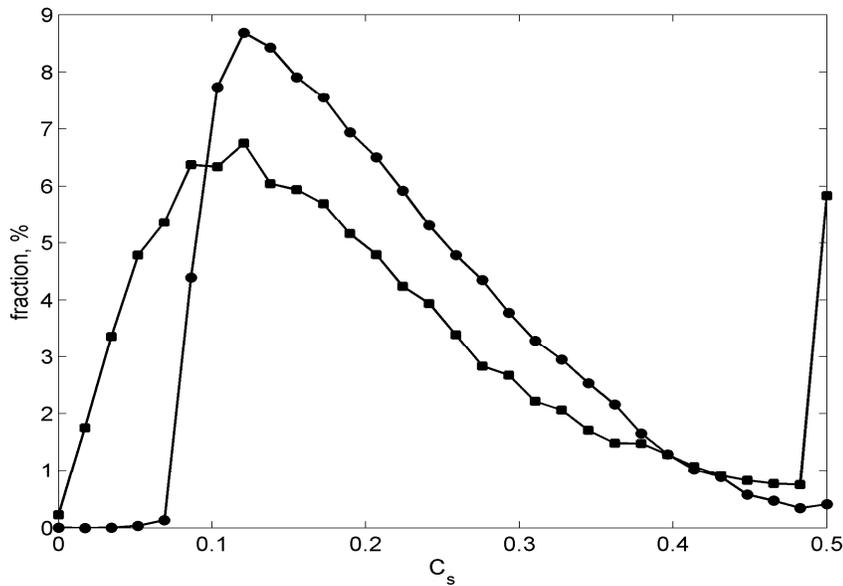

**Figure 4.** Distribution of $C_s$ at the levels 1 through 6 in the LES without (squares) and with ANN (dots) correction.

*b. Artificial neural network in the sub-grid scale modelling*

A correct way to introduce the unresolved boundary conditions into the LES model is through the sub-grid scale model. The sub-grid scale model describes the integral effect of unresolved flow features on the resolved scales. Sarghini et al. [22] pioneered application of ANN to dynamic sub-grid scale modelling. LES equations are obtained through filtering of Navier-Stokes equations. The filtering result in appearance of a sub-grid stress term

$$\tau_{ij} = \overline{u_i u_j}^h - \overline{u_i}^h \overline{u_j}^h - 2l_{cut}^2 |\overline{S}_{ij}^h| \overline{S}_{ij}^h, \tag{5}$$

which must be modelled. For more details see Esau [3]. Here the overbar with the superscript $h$ denotes grid filtering; $l_{cut} = C_s \Delta$ is the mixing length scale; $C_s$ is the Smagorinsky constant; $\Delta$ is the LES mesh resolution. The strain rate tensor is

$$S_{ij} = \frac{1}{2}\left(\nabla_i u_j + (\nabla_j u_i)^T\right), \tag{6}$$

where $\nabla_i$ denotes a gradient in *i*-th direction and superscript $T$ denotes a conjugate transposed.

As Eq. (5) reveals $C_s$ is the parameter to be tuned to the specific unresolved features of the flow. Here we consider the following numerical experiments to test the ANN utility in the described problem of the flow around large unresolved obstacles below the first computational layer depicted in Figure 3. The numerical experiment with the LESNIC model simulates with developed atmospheric turbulent boundary layer with the geostrophic wind speed 4 m s$^{-1}$ at latitude 45 degrees over homogeneous flat surface with the roughness length scale 0.01 m. The model domain was 1000 m by 750 m by 500 m in streamwise, spanwise and vertical directions with the mesh 64 by 48 by 64 grid nodes. The boundary layer evolved over one hour of model time. At the end of its evolution, the boundary layer occupied ½ of the domain depth. The last 3-dimensional state of the model was stored and later used as initial conditions to obtain additional 30 min runs both with homogeneous and heterogeneous surfaces. The simulations were run for truly neutral boundary layers with passive scalars without feedback effects on the flow. The lateral boundary conditions were periodic in all runs. At the surface, the turbulent flux of momentum was computed instantly and pointwise using the log-law and therefore did not account directly to the surface features.

The ANN was utilized in the following way. The 3 layer backpropagation feeding ANN was trained on 5 min data from the 30 min additional LES runs described above. The input vector for the ANN was composed of components of the tensor in Eq. (6) and its module as

$$X = [S_{11}, S_{12}, S_{13}, S_{21}, S_{22}, S_{23}, S_{31}, S_{32}, S_{32}, |S_{ij}|]^T. \tag{7}$$

Since the tensor is symmetric, its off-diagonal components receive larger weights in the input vector. Let us first consider the effect of the ANN in the Ekman layer simulations over flat, homogeneous surface. Figure 4 compares histograms of $C_s$ computed with Eq. (3) only and with Eq. (3) following the ANN correction. Pleasant modifications introduced into the $C_s$ distribution by ANN are clearly seen. The small values of $C_s$, which destabilize the simulations, were almost eliminated and too large values ($C_s > 0.4$) occurred considerably less frequent in the flow during the analysed 30 min interval of time. Thus, we can conclude that sub-grid modelling with ANN correction may improve the performance and stability of LES even in simulations over homogeneous surfaces.

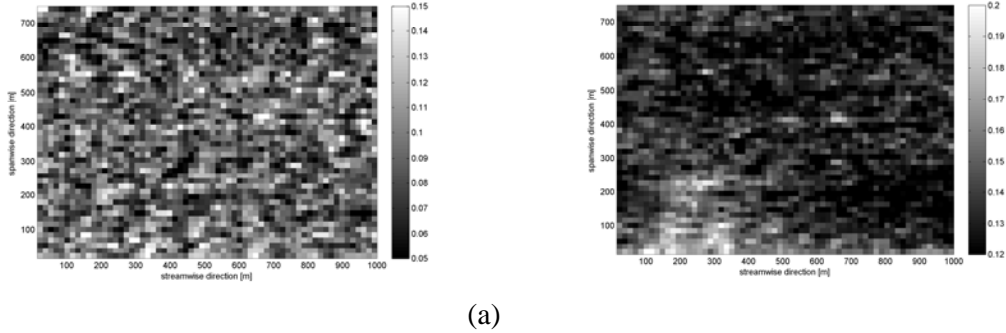

(a) (b)

**Figure 5.** Horizontal map of $C_s$ at the 1st level in the coarse resolution LES without (a) and with ANN (b) correction. Larger $C_s$ (light areas) corresponds to larger sub-grid fluxes in response to the resolved velocity shear.

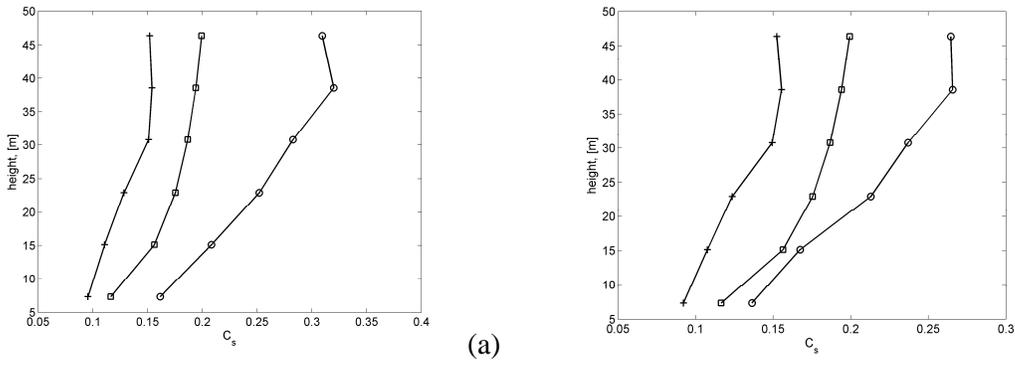

(a) (b)

**Figure 6.** Averaged vertical profiles of Smagorinsky coefficient, $C_s$, in the downstream (a) and aside (b) sub-domains in LES-H1 experiment: crosses – dynamically calculated coefficient; circles – ANN modified coefficient; squares – ad hoc fitting from Balaras et al. (1995).

**Table 1.** Domain-averaged correlation between actual, i.e. simulated in dynamic LES, sub-grid turbulent stress in the high resolution run and the stress in the control run with the ANN correction using the input vector *X* over post training period of 25 min.

| Experiment | Correlation | Comments |
|---|---|---|
| Homogeneous surface | 0.585 | Low correlation due to filtering of too low $C_s$ values in ANN; fraction of $C_s < 0.05$ is 48%. |
| Heterogeneous surface roughness, LES-H1 | 0.595 | Domain-averaged correlation improves as there is generally higher $C_s$. Fraction of $C_s < 0.05$ is 40%. |
| Heterogeneous surface with an obstacle, LES-H2 | 0.581 | Domain-averaged correlation improves as there is generally higher $C_s$. Fraction of $C_s < 0.05$ is 39%. |

The next step is to introduce a surface heterogeneity in form of rough spots that represent completely unresolved obstacles (e.g. **C** in Figure 3) or just surface heterogeneity of any other kind. This experiment is LES H1 in Table 1. The surface properties do not change during the simulations. Hence ANN can be trained to correct the simulations specific features of the perturbed flow. This training is achieved in the study trough the run with twice as fine resolution, which served as observations for the training process. The run LES-H1 had only surface roughness modified. The roughness was set to 5 m in a square of ¼ of the domain surface area; whereas the rest of the domain had roughness of 0.01 m. Figure 5 shows the mean values of $C_s$ in the coarse resolution run without ANN (Figure 5a) and with ANN (Figure 5b) correction. The added value of the ANN correction is clearly seen in localized modification of $C_s$ and therefore the surface sub-grid fluxes as related in Eq. (5). Figure 6 shows that ANN was able to improve the vertical profile of averaged $C_s$ as well. ANN has automatically introduced modifications, which were previously left for *ad hoc* fitting as described in Balaras et al. [24]. The observed changes in the mean Smagorinsky coefficient profiles as well as in their spatial distribution and clipping of the extreme values are welcome modifications made through ANN implementation to LES closure problem over roughness heterogeneous surface.

Now, we consider a more complicated case schematically described in Figure 3 with a rectangular obstacle **B.** The obstacle was inserted explicitly into the fine resolution LES run, whereas in the coarse resolution run LES-H2 it is under-resolved in the sense that only the 1$^{st}$ model level was modified to have an immersed boundary conditions. The log-law was applied only at the upward facing facet of the obstacle, whereas the simple non-slip conditions were applied on the lateral facets. The obstacle in LES-H2 is of 12 m tall (2 computational levels) located in ¼ of the domain area. Hence, LES-H1 represented a resolved flow response on unresolved elements of urban morphology, and LES-H2 represents a corresponding response on semi-resolved elements. Figure 7 shows values of $C_s$ obtained as in Figure 5 but for LES-H2 experiment. An impenetrable obstacle indeed has strong impact on $C_s$ values and their spatial distribution. As in Glazunov [25], $C_s$ is distributed asymmetrically around the obstacle with larger values in the front (upstream) facet and smaller values at the rear (downstream) facet. Remarkably, ANN correction was able to account for the localization of the smallest $C_s$ on the downstream sides of the obstacle and very close to the centre of the obstacle front facet. Those features were obtained in a fine resolution high quality LES by Glazunov [25] but hardly reproduced by the dynamic closure in this LES. It is also worth noticing that the time needed to estimate $C_s$ with the ANN correction is about 2 orders of magnitude less than time used in the Glazunov's LES (personal communication) and an order of magnitude less than time used to compute it in the present high-resolution LES.

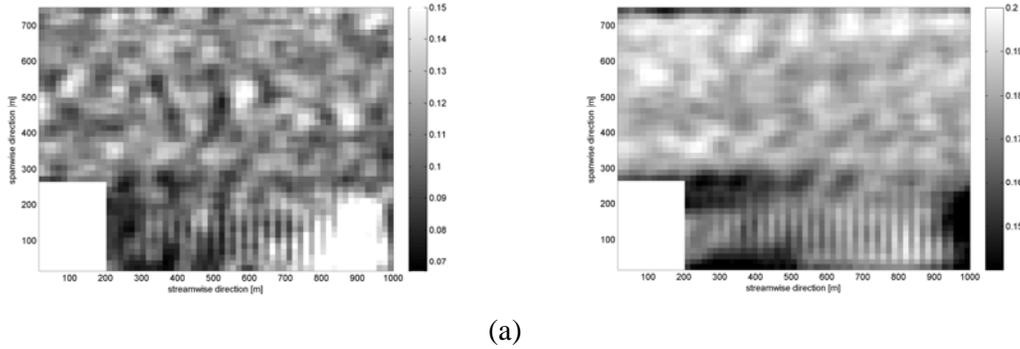

(a)                                                                      (b)

**Figure 7.** The same as in Figure 5 but for LES-H2 experiment. Pay attention that in Figure 5 the turbulent fluxes are computed over the roughness spot itself, whereas in this experiment the flow does not penetrate into the obstacle.

*c. Artificial neural network in modelling of concentrations*

One of the most important and interesting applications of ANN in conjunction with LES is the modelling of the scalar concentrations (pollution). The problem is to predict the concentration field $q(\vec{x},t)$ on the basis of the velocity field $\vec{u}(\vec{x},t)$, which is known with a limited accuracy, and $q(x_n, t-n\delta t)$, which is know only in a few locations $x_n$. Using our control LES run, we computed 1 min forecast of concentrations in the near field of a pollution source in the turbulent boundary layer. The forecast was computed 25 times over 25 min of model time. Initially, ANN has been trained on 5 min data in order to produce $q(\vec{x},t)$. The input vector was composed of components of velocity and the historical value $q(x_n, t-\delta t)$ where $\delta t = 1$ min. Hence, the input vector was defined as $X = [u_1(t), u_2(t), u_3(t), q(t-\delta t)]^T$. Figure 8 shows concentration field averaged over 30 min and within the layer $\pm 15$ m above and below the injection source. The source was located at the height of 18 m above the surface so that the polluted wake meanders in three dimensions. The spots of high concentration are created by the wake meandering in the vertical direction. A nontrivial feature is a rarefied scalar concentration directly downstream from the source on the wake axes recognized also in Babatunde and Enger [26]. ANN predicts robustly short-term fluctuations of the concentration but produce some noise in the absence of a real signal. The overall weighted correlation between the simulated concentration and the ANN averaged forecast was 0.92 in this experiment. The forecast interval of 1 min would probably seem too short for a practical use but the reader should keep in mind that we are analysing ANN as an addition to a hydro-dynamical model or LES. Thus, the ANN role is to predict only on unresolved intervals of time. The advection across larger distances and longer time scales are explicitly resolved in the model. Indeed over 1 min even with the wind speed of 4 m s$^{-1}$ a polluted air volume would be transported over 240 m or more than 10 grid intervals in the present experiment. With typical wind of 10 m s$^{-1}$, the same volume will be transported over 600 m or 50 grid intervals.

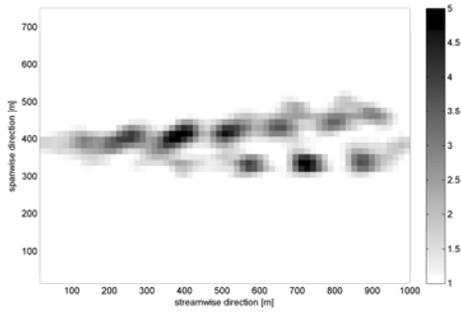 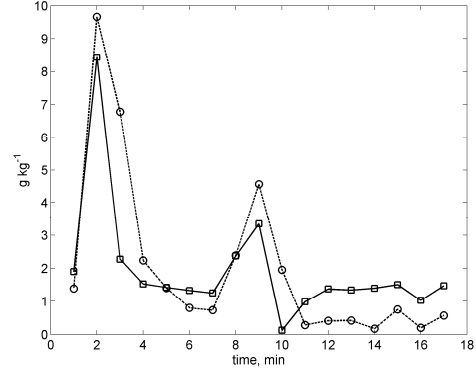

(a) (b)

**Figure 8.** (a) The averaged over 30 min field of concentration (g kg$^{-1}$) of a passive scalar released from unresolved source at 18 m above the surface. (b) The sequence of the 1 min concentration forecasts with ANN (squares) using pollution data from homogeneous LES experiment (circles).

*d. Discussion: Compatibility of LES and ANN approaches*

In what was presented above, the concentration at un-resolved scales was treated as proportional to the resolved-scale concentration with sub-grid scale fluxes equal to those for momentum with the constant Schmidt number of unity. Rigorous LES formalism, exemplified in Cook and Riley [27] model, demands to express the sub-grid quantity $q'$ through resolved quantity $\bar{q}$ in a universal scale-similar manner. Cook and Riley exploited the fact that any quantity defined at [0;1], i.e. normalized $q$, and subjected to stochastic turbulent fluctuations $u$ with the Gaussian distribution function should have the $\beta$-distribution

$$\beta(x;\bar{q},\sigma_q^2) = x^{a-1}(1-x)^{b-1}\frac{\Gamma(a+b)}{\Gamma(a)\Gamma(b)}, \qquad (8)$$

where $\Gamma(a)$ is the gamma function of Euler and the coefficients are

$$a = \bar{q}\left(\frac{\bar{q}(1-\bar{q})}{\sigma_q^2}-1\right), b = \frac{a}{\bar{q}} - a. \qquad (9)$$

With knowledge of the $\beta$-distribution, the true values of concentration and its ensemble-averaged value over a grid cell are related as (Jimenez et al. [28])

$$<q> = \int_\Delta q(x)\beta(x;\bar{q},\sigma_q^2)dx. \qquad (10)$$

In principle, Eqs. (8)-(10) allow estimations of the concentration within the model grid cell with a minimal root mean square error. It is worth mentioning that for air quality analysis, it is also interesting to estimate extreme concentrations. Cook and Riley approach is not suitable for it. The scalar dispersion $\sigma_q^2$ is unknown but can be determined from the resolved scalar concentrations at relatively low Reynolds number as

$$\sigma_q^2 \propto \overline{\tilde{q}^2} - \tilde{\bar{q}}^2 , \tag{11}$$

where tilde denote another somewhat larger filter. As Jimenez et al. [28] argued, the envisaged model provides excellent results for homogeneous flow even at high Re. The situation is more complicated in heterogeneous urban sub-layer. Here, space-time filtering applied to compute $\sigma_q^2$ and ultimately $<q>$ is not valid any more since the filter stencil would cross the walls. Moreover, the double filtering is not very useful as well. The resolved variations of the concentration are far too small to represent scale similarity. Recently Tseng et al. [29] have reported some degree of success in simulations of the air quality over Baltimore downtown with LES with Lagrangian scale depended sub-grid model, which use filtering alone a flow trajectory instead of spatial filtering.

## 4. Conclusions

The micro-meteorological modelling tool, large-eddy simulation model, is well established for the research purposes. In order to utilize this tool in practical applications, two essential components must be developed. They are: (a) the data assimilation on meteorological micro-scales, where the effects of the turbulence are not negligible; (b) the parameterization of the unresolved surface and emission features with effect on the resolved turbulence. At present, the LES is of limited utility for applications as it cannot start from (i) scarcely measured atmospheric conditions and (ii) it cannot account for unresolved surface details. This study presented an analysis of several numerical experiments with the LESNIC LES code. The study addressed two abovementioned problems.

The numerical experiments demonstrated that the simplest nudging or Newtonian relaxation technique for the data assimilation is applicable on the turbulence scales. It is reasonable to claim that the nudging technique is a promising candidate for the data assimilation in the turbulence-resolving LES. The study does not reveal any significant distortions of the turbulent spectra of the total turbulent energy of fluctuations on the essential scales of the turbulence generation and down to the dissipation. Even in the case of dynamical inconsistency of the observed profile, the LES tries to generate proper turbulent fluctuations through intensification of the turbulence in one or another sub-layer where the stability restrictions are relaxed.

We considered a possibility to combine LES and ANN approaches to deal with complex surface features and unresolved emission sources. Here, *a priori* analysis has been applied for idealized, relatively coarse resolution numerical experiments, namely, the experiments with turbulent neutrally stratified boundary layers over homogeneous surface, the surface with very large roughness patch and the surface with an obstacle. It is shown that the filtering property of the three layers' ANN is useful in improvement of numerical stability and physical consistency of the dynamic sub-grid closure in LES. The 3-layer tan-sigmoid ANN with back-propagation training over rather short interval of 5 min outperformed the usual engineering clipping approach to obtain the dynamic values of the Smagorinsky constant. ANN for each of dynamical sub-regions in the LES domain could greatly reduce computer time needs. ANN is shown to be a robust predictor for the pollutant scalar concentration in the urban sub-layer with unresolved scalar sources.

**Acknowledgements**

Discussions with Dr. A. Baklanov, Dr. M. C. Mammarella and ATMOSFERA group members G. Grandoni and P. Fedele were inspiring. The work was partially supported by EU FP6 Networking Project *MODOBS – exploring added values from new modeling and observation techniques;* EU FP7 collaborative project MEGACITIES – *Emissions, urban, regional and Global Atmospheric POLlution and climate effects, and Integrated tools for assessment and mitigation;* Norwegian Research Council grant *180343/S50 Analysis and Possibility for Control of Atmospheric Boundary Layer Processes to Facilitate Adaptation to Environmental Changes*.
**References**

[1] Wyngaard, J.C., "Toward Numerical Modeling in the "Terra Incognita". *J. Atmos. Sci.*, vol. **61**, pp. 1816–1826, 2004

[2] Esau, I., "Large-eddy simulations of geophysical turbulent flows with applications to planetary boundary layer research", arXiv:0907.0103v1, 2009

[3] Esau, I., "Simulation of Ekman boundary layers by large eddy model with dynamic mixed subfilter closure", *J. Environ. Fluid Mech.*, vol. **4**, pp. 273-303, 2004

[4] Porson, A., J. Price and A. Lock, "Modelling of observed fog development over flat terrain", *Geophys. Res. Abs.*, vol. **11**, EGU2009-3155, 2009

[5] Cheng, W. Y. Y., Wu, T. and W.R. Cotton, "Large-eddy simulations of the 26 November 1991 FIRE II Cirrus Case", *J. Atmos. Sci.*, vol. **58**, pp. 1017-1034, 2001

[6] Carrio G., Cotton, W.R. and D. Zupanski, "Data assimilation into a LES model: Retrieval of IFN and CCN concentrations", AGU meeting, 2005

[7] Losch, M. and C. Wunsch, "Bottom Topography as a Control Variable in an Ocean Model", *J. Atmos. Oceanic Technol.*, vol. **20**, pp. 1685–1696, 2003

[8] Kazantsev E., "Identification of an optimal boundary approximation by variational data assimilation", *J. Comp. Phys.*, 2009, in print

[9] White, H., "Connectionist Nonparametric Regression: Multilayer Feedforward Networks Can Learn Arbitrary Mappings", *Neural Networks*, vol. **3**, pp. 535-550, 1990

[10] Lee, C., Kim, J., Babcock, D. and R. Goodman, "Application of neural networks to turbulence control for drag reduction", *Phys. Fluids*, vol. **9**, n. 6, pp. 1740 -1751, 1997

[11] Ayhan, T., Karlik, B. and A. Tandiroglu, "Flow geometry optimization of channels with baffles using neural networks and second law of thermodynamics", *Comput. Mech.*, vol. **33**, pp. 139–143, 2004

[12] Hocevar, M., Sirok, B. and I. Grabec, "A Turbulent-Wake Estimation Using Radial Basis Function Neural Networks", *Flow Turbul. Combustion*, vol. **74**, pp. 291–308, 2004

[13] Panigrahi, P. K., Dwivedi, M., Khandelwal, V. and M. Sen, "Prediction of Turbulence Statistics Behind a Square Cylinder Using Neural Networks and Fuzzy Logic", *J. Fluids Engineering*, vol. **125**, n. 2, pp. 385-387, 2003

[14] Morabito, F. C. and M. Versaci, "Wavelet neural network processing of urban air pollution", *Proceedings of International Joint Conference on Neural Networks*, vol. **1**, pp. 432 – 437, 2002

[15] Zhang, Z. and Y. San, "Adaptive wavelet neural network for prediction of hourly NOx and NO2 concentrations", *Proceedings of Winter Simulation Conference*, eds. Ingalls et al., pp. 1770-1778, 2004

[16] Kukkonen, J. and 10 co-authors, "Extensive evaluation of neural network models for the prediction of NO2 andPM 10 concentrations, compared with a deterministic modelling

system and measurements in central Helsinki", *Atmos. Environ.*, vol. **37**, pp. 4539–4550, 2003

[17] Mammarella, M. C., Grandoni, G., Fedele, P., Sanarico, M. and R.A. Di Marco, "Neural networks for predicting and monitoring urban air pollution: the ATMOSFERA intelligent automatic system", *Accademia dei Lincei*, 2006

[18] Von Storch, H., Langenberg, H. and F. Feser, "A spectral nudging technique for dynamical downscaling purposes", *Mon. Wea. Rev.*, vol. **128**, pp. 3664-3673, 2000

[19] Andren, A., Brown, A., Graf, J., Mason, P., Moeng, C.-H., Nieuwstadt, F.T.M. and U. Schumann, "Large-eddy simulation of a neutrally stratified boundary layer: A comparison of four computer codes", *Quart. J. Roy. Meteorol. Soc.*, vol. **120**, pp. 1457–1484, 1994

[20] Mauritsen, T., Svensson, G., Zilitinkevich, S. S., Esau, I., Enger, L., and B. Grisogono, "A total turbulent energy closure model for neutral and stably stratified atmospheric boundary layers", *J. Atmos. Sci.*, vol. **64**, n. 11, pp. 4117-4130, 2007

[21] Moreau, A., Teytaud, O. and J. P. Bertoglio, "Optimal estimation for large-eddy simulation of turbulence and application to the analysis of sub-grid models", *Phys. Fluids*, vol. **18**, n. 10, pp. 10510-10511, 2006

[22] Sarghini, F., de Felice, G. and S. Santini, "Neural networks based sub-grid scale modeling in large eddy simulations", *Computers & Fluids*, vol. **32**, pp. 97–108, 2003

[23] Hsieh, W.W. and B. Tang, "Applying neural network models to prediction and data analysis in meteorology and oceanography", *Bull. Am. Meteorol. Soc.*, vol. **79**, pp. 1855-1870, 1998

[24] Balaras, E., Benocci, C. and U. Piomelli, "Finite-difference computations of high Reynolds number flows using the dynamic subgrid-scale model", *J. Theor. Comput. Fluid Dynamics*, vol. **7**, n. 3, pp. 207-216, 1995

[25] Glazunov, A., "Large eddy simulation of rough-wall-bounded turbulent channel flow using localized dynamic closure and high-order numerical scheme", *Geophys. Res. Abstracts*, vol. **9**, 09088, 2007

[26] Babatunde, A. J. and L. Enger, "The role of advection of fluxes in modelling dispersion in convective boundary layers", *Quart. J. Roy. Meteorol. Soc.*, vol. **128**, n. 583, pp. 1589-1607, 2002

[27] Cook, A. W. and J. J. Riley, "A sub-grid model for equilibrium chemistry turbulent flows", *Phys. Fluids*, vol. **6**, pp. 2868, 1994

[28] Jimenez, J., Linan, A., Rogers, M. M. and F. J. Higuera, "A priori testing of sub-grid models for chemically reacting non-premixed turbulent shear flows", *J. Fluid Mech.*, vol. **349**, pp. 149-171, 1997

[29] Tseng, Y.-H., Meneveau, C. and M. Parlange, "Modeling flow around bluff bodies and predicting urban dispersion using large eddy simulation", *Environ. Sci. Technol.*, vol. **40**, pp. 2653-2662, 2006